\documentclass[preprint]{imsart}

\usepackage{algorithm}
\usepackage{algpseudocode}
\usepackage{graphicx}
\usepackage{verbatim}
\usepackage[sumlimits,]{amsmath}
\usepackage{amsfonts}
\usepackage[numbers]{natbib}
\RequirePackage[OT1]{fontenc}
\RequirePackage{amsthm,amsmath,amssymb}
\RequirePackage[colorlinks,citecolor=blue,urlcolor=blue]{hyperref}

\def \d { {\,\mbox{d}}}

\def \dt { \,\mbox{d}t}

\def \dx { \, \mbox{d}x}

\def \as { {\emph{a.s.}\,}}

\def \calN { {\mathcal N}}

\def \calU { \mathcal{U}}

\def \calY { \mathcal{Y}}

\def \eps { {\varepsilon}}

\def \st {:\,}

\def \iid { {\mbox{i.i.d.}\,}}
\def \textiid {i.i.d.\ }

\def \E { {\mathbb E}}

\def \p {\partial}

\newcommand{\Exp}[1]{ \E\left\{ #1 \right\} }
\newcommand{\Exparg}[2]{ \E_{#1}\left\{ #2 \right\} }

\newcommand{\KL}[2]{ \mbox{KL}[#1 \,||\, #2]}   
\newcommand{\qbox}[1] { {\quad\mbox{#1}\quad}}
\newcommand{\Trace}[1] { {\mbox{Trace}\left\{ #1 \right\}}}
\newcommand{\Var}[1]{ \mbox{Var}\left\{ #1 \right\} }

\def \Nat {{\mathbb N}}
\def \Rone { {\mathbb R}}

\def \RN {{\mathbb R}^N}

\newcommand{\pushfwd}[1]{ {#1_{\#}}}

\newcommand{\Wishart}[2]{ Wishart(#1, #2) }
\newcommand{\InverseWishart}[2]{ InverseWishart(#1, #2) }
\def \Chat { {\hat{C}}}
\def \Lhat { {\hat{L}}}

\def \W  { {\kappa}}

\startlocaldefs
\numberwithin{equation}{section}
\theoremstyle{plain}
\newtheorem{theorem}{Theorem}[section]
\newtheorem{lemma}{Lemma}[section]
\newtheorem{corollary}{Corollary}[section]
\newtheorem{proposition}{Proposition}[section]
\theoremstyle{remark}

\newtheorem{desiderata}{Desiderata}[section]
\endlocaldefs

\begin{document}

\begin{frontmatter}

\title{A Condition Number for Hamiltonian Monte Carlo}
\runtitle{A Condition Number for HMC}

\begin{aug}
\author{\fnms{Ian Langmore} \thanksref{e1} \ead[label=e1,mark] {ianlangmore@gmail.com}},
\author{\fnms{Michael Dikovsky}},
\author{\fnms{Scott Geraedts}},
\author{\fnms{Peter Norgaard}}, \and
\author{\fnms{Rob von Behren}}
\address{Google LLC. 1600 Amphitheatre Parkway. Mountain View, CA. 94030, \printead{e1}}

\affiliation{Google Research}
\runauthor{I. Langmore et al.}

\end{aug}

\maketitle

\begin{abstract}
Hamiltonian Monte Carlo is a popular sampling technique for smooth target densities.  
The scale lengths of the target have long been known to influence integration error and sampling efficiency.  However, quantitative measures intrinsic to the target have been lacking.
In this paper, we restrict attention to the multivariate Gaussian and the leapfrog integrator, and obtain a condition number corresponding to sampling efficiency.  This number, based on the spectral and Schatten norms, quantifies the number of leapfrog steps needed to efficiently sample.  We demonstrate its utility by using this condition number to analyze HMC preconditioning techniques.  We also find the condition number of large inverse Wishart matrices, from which we derive burn-in heuristics.
\end{abstract}

\begin{keyword}
  \kwd{Hamiltonian Monte Carlo}
  \kwd{Markov Chain Monte Carlo}
  \kwd{Condition Number}
  \kwd{Preconditioning}
  \kwd{Random Matrices}
\end{keyword}

\end{frontmatter}

\section{Introduction}
\label{section:introduction}
Hamiltonian Monte Carlo (HMC) is a technique for sampling random variables $X\in\RN$, possessing smooth densities $p(x)$.  A a core step is the numerical integration of Hamilton's equations for time $T$, in $\ell$ discrete steps, each of size $h$.
Tools are available to adjust $h$ and $\ell$ so as to maximize sampling efficiency for particular problems \citep{Lao2020-qi,Carpenter2017-zd,Andrieu2008-xv}.  Lacking has been a measure of difficulty, intrinsic to the density $p(x)$, rather than sub-optimal choices of $h$ and $\ell$.

It has long been recognized that disparate covariance scales in $X$ tend to make sampling difficult \citep{Betancourt2017-od}.  This motivates techniques to ``flatten'' $X$ through transformations. These transformations can be as basic as scaling components of $X$ by their standard deviation, or as complex as application of a diffeomorphism built with polynomials or a neural network \citep{Parno2018-ch,Hoffman2019-pd}.  Despite some success, there is limited understanding as to exactly \emph{how much} better or worse different covariance scales are.

Our main contribution is to show that, in the multivariate Gaussian case, one particular condition number governs the number of leapfrog integration steps needed to efficiently sample in every direction.  This number, $\W$ (see \eqref{align:kappa-condition-number}), differs from the common spectral condition number (ratio of largest to smallest singular values) since it takes into account all eigenvalues of the covariance matrix.  This is needed, since all eigenvalues contribute to integration error.

Using $\W$ we are able to analyze and develop preconditioning techniques.
We find the law of $\W$ when preconditioning with the sample covariance. The law turns out to be the condition number of the inverse Wishart ensemble.
An asymptotic expression for this law is then derived, leading to a set of preconditioning heuristics.  We next show that the popular component-wise standardization can be better or worse than preconditioning with a diagonal transformation trained via variational inference.  Each of these in turn can be better or worse than doing nothing at all.  Just as importantly, insight is given into what sort of spectra are antithetical to efficient HMC.  These ``bad'' spectra have only a few large eigenvalues, and many small ones.  
We demonstrate the virtue of a low rank update preconditioner for this situation.

We limit analysis to Gaussian targets, despite the fact that sampling from them does not even require HMC.
Our hope is that, by providing \emph{explicit} formulas and \emph{precise} analysis, these results will be helpful in more general situations.
For example, the original motivation for this work was to determine if reverse or forward KL preconditioning was strictly superior (the answer is ``No'').  Additionally, a pleasant consequence of a condition number formulation is the connection to random matrix theory.  We hope to expand on this line of thinking in future work.


Our result relating step size to integration error (theorem \ref{theorem:integration-error}) is similar to the analysis of \citep{Beskos2013-nd,Betancourt2014-kg}, which derive scaling laws for (non-Gaussian) densities with repeated components.  Our restriction to Gaussian densities allows for an explicit relationship between scales of the random variable $X$ and the necessary step size/number of integration steps, and removes the repeated components requirement.  A condition number is used for selecting preconditioners in \citep{Bales2019-fz}.  Their number shows some promise for non-Gaussian systems, but does not have a precise relation to sampling efficiency (our $\W$ does).

In section \ref{section:hmc-basics} we briefly review the HMC method.  Section \ref{section:hmc-work} goes over our main results surrounding our condition number $\W$.  Section \ref{section:kappa-for-inv-wishart} derives an asymptotic result on $\W$ for the inverse Wishart ensemble.  Section \ref{section:preconditioning} demonstrates using $\W$ to derive and analyze different preconditioning techniques.  Proofs of the main results are in section \ref{section:proofs}.

\section{The Hamiltonian Monte Carlo Method}
\label{section:hmc-basics}
Here we quickly review the basics of HMC for purposes of establishing notation.  A comprehensive introduction can be found in \citep{Neal2012-jd}.

The Hamiltonian Monte Carlo (HMC) method was introduced in 1987 as ``Hybrid Monte Carlo'' for use in lattice field theory simulations \citep{Duane1987-eg}.  Since then, it has been recognized as an efficient alternative to random walk Metropolis, well suited for higher dimensional problems.  Implementations are available for a variety of languages \citep{Carpenter2017-zd}.

HMC defines a way to sample from smooth densities $p(x)$ for $X\in\RN$ by augmenting state space with a momentum $\xi\in\RN$, and defining the joint density
\begin{align*}
  p(x,\xi) &= \exp\left\{ -H(x,\xi) \right\},\quad\mbox{where}\quad H(x,\xi) := - \log p(x) + \frac{\|\xi\|^2}{2},
\end{align*}
where $\|\xi\|$ is the Euclidean norm.  Alternative norms may be used, although these are less popular in practice \citep{Girolami2011-wv}.  Moreover, a fixed norm generated through the inner product $\langle LL^T\xi,\xi\rangle$ is shown in \citep{Neal2012-jd} to be equivalent to the linear preconditioning $X\mapsto LX$ (which we \emph{do} consider here).

In the physics setting, the \emph{Hamiltonian} $H$, is total energy, whereas $-\log p(x)$, $\|\xi\|^2/2$ are potential and kinetic energies.  Sampling proceeds by (a numerical approximation to) the following iteration from point $(x^j, \xi^j)$.
\begin{enumerate}
  \item Draw $\tilde \xi\sim\calN(0, I_N)$.
  \item Let $(x(t), \xi(t))$ be the time $t$ solution to the ODE $\dot x = \xi$, $\dot \xi = \nabla \log p(x)$, with initial condition $(x^j, \tilde\xi)$.
  \item Set $(x^{j+1}, \xi^{j+1}) = (x(T), -\xi(T))$, for integration time $T$.
\end{enumerate}

In practice, the ODE must be solved numerically over $\ell$ steps with step-size $h$. Denote this solution by $\Psi^\ell$.  The integration error means we can no longer just accept the move in step 3, which is replaced by a Metropolis correction:
\begin{align*}
  (x^{j+1}, \xi^{j+1}) &= \Psi^\ell,\quad\mbox{with probability } a( x^j, \xi^j \to \Psi^\ell),
\end{align*}
and
\begin{align*}
  (x^{j+1}, \xi^{j+1}) &= (x^j, \xi^j),\quad\mbox{with probability } 1 - a( x^j, \xi^j \to \Psi^\ell),
\end{align*}
for acceptance probability
\begin{align}
  \label{align:metropolis-step}
  a(x^j, \xi^j \to \Psi^\ell)
  :&= \min\left( 1, \exp\left\{ H(x^j, \xi^j) - H(\Psi^\ell) \right\} \right).
\end{align}
Since Hamilton's equations of motion preserve the Hamiltonian, if numerical integration was perfect, $H(x^j, \xi^j) = H(\Psi^\ell)$ and every step would be accepted.  In practice, finite step size leads to some rejections and wasted effort.

\begin{figure}
  \centering
  \includegraphics[width=0.98\textwidth]{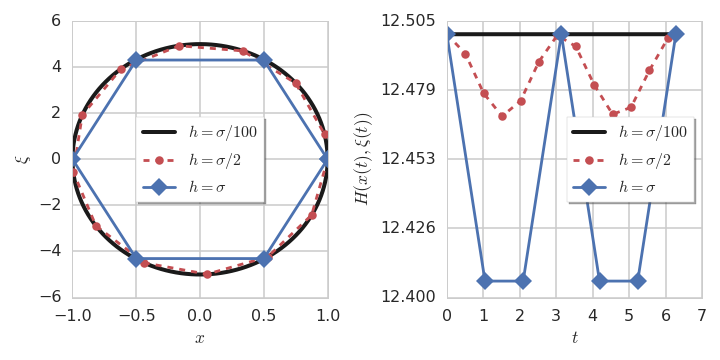}
  \caption{\textbf{Leapfrog Integration Error}.
  \textbf{Left:} Integrating trajectories $(x(t), \xi(t))$ with different step size $h$.  $h=\sigma/100$ is nearly perfect.  Even with larger $h$, deviation of trajectories from the perfect line are barely perceptible.
  \textbf{Right:} Values of the Hamiltonian, $H(x(t), \xi(t))$, along the trajectories.  For larger $h$, the error is greater, but does not diverge.}
  \label{fig:integration}
\end{figure}

The numerical integration is usually done with $\ell$ steps of the \emph{St\"ormer-Verlet} or \emph{leapfrog integrator}, each step progressing $(x, \xi)$ to $(x_h, \xi_h)$ via
\begin{enumerate}
  \item Set $\xi_{h/2} = \xi + \frac{h}{2}\nabla\log p(x)$
  \item Set $x_h = x + h \xi_{h/2}$
  \item Set $\xi_h = \xi_{h/2} + \frac{h}{2}\nabla\log p(x_h)$
\end{enumerate}
Figure \ref{fig:integration} shows that integration errors remain small, even if $h \approx \sigma$.  Just as importantly, trajectories do not diverge, but instead follow paths of a modified Hamiltonian due to the fact that the leapfrog integrator is symplectic \citep{Neal2012-jd,Leimkuhler2005-zb}.

The number of leapfrog steps $\ell$ is often chosen to be a fixed (but highly influential) constant.  To avoid unlucky (or difficult to analyze) circumstances, we use a random integration time $T$, then set $\ell = \lceil T/h\rceil$.  This randomness is often introduced to ensure ergodicity.  See section 3.2 of \citep{Neal2012-jd} as well as \citep{Mackenze1989-ww}.  Inspection of our proofs show (see e.g. \eqref{align:Delta-N-mean}), without this regularizing effect the spectrum \emph{could} conspire to make the leading term vanish.  Additionally, with a fixed integration length and dense enough spectrum, near \emph{resonances} can occur, whereby samples nearly repeat the same trajectory.

\section{$\W$ and Computational Effort in HMC}
\label{section:hmc-work}
Important results relating computational effort and step size has been established by previous work:
Being a second order integrator, leapfrog results in error in the Hamiltonian, bounded at each step by $O(h^3)$. Thus, over $\ell = T / h$ steps, error is bounded by $O(T h^2)$ \citep{Neal2012-jd,Leimkuhler2005-zb}.  In expectation the situation is better: \citep{Beskos2013-nd,Betancourt2014-kg} show that asymptotically, the integration error is Normal with scale $O(h^2)$, with $T$ contributing only to higher order terms.

In this section, we establish a relationship between the covariance spectrum, and the number of leapfrog steps needed to effectively sample.  This is rigorously analyzed as dimension $N\to\infty$.  Before submersing into the world of limits, consider fixed dimension and covariance spectrum $\sigma_1^2\geq\sigma_2^2\geq\cdots\geq\sigma_N^2>0$.
Our results are motivated by a practitioner who adjusts the step size $h$ and number of leapfrog steps $\ell$ to achieve the following
\begin{desiderata}
  \label{desiderata:hmc}
  (i) The average acceptance probability $\Exp{a(X, \xi \to \Psi^\ell)} = \bar a$, for desired $\bar a \in (0, 1)$.
  (ii) The number of integration steps $\ell = \lceil \sigma_1 T/ h\rceil$, for integration time $T\sim\pi$, where $\pi$ is some probability density.
\end{desiderata}
To motivate (i), consider results in \citep{Beskos2013-nd, Betancourt2014-kg}, where computational cost is shown to be (asymptotically in dimension) optimal when the average acceptance probability approaches a limit (approximately 0.68).  In addition to being asymptotically optimal, tuning $h$ to achieve desired $\bar a$ is often convenient \citep{Andrieu2008-xv,Carpenter2017-zd}.
Condition (ii) states that each trajectory travels a distance $h\ell$ that scales with the largest scale length, $\sigma_1$. This prevents HMC from reverting to a random walk in the direction corresponding to $\sigma_1$.

Condition (ii) does imply $\ell$ could be quite large, but this turns out not to be an issue. The reason, is that the dominant error term depends on integration length only through an average of $\sin^2(\cdot)$, which is close to being constant (see the proof of theorem \ref{theorem:integration-error} in section \ref{section:proofs:normal-limit}).  Therefore, the acceptance rate depends strongly on $h$ but only weakly on $\ell$. Thus, the user can often adjust $\ell$ \emph{after} setting $h$, so that $h\ell\propto \sigma_1$ as desired.

Define
\begin{align}
  \label{align:WN}
  \W:= \left( \sum_{n=1}^N \left(\frac{\sigma_1}{\sigma_n}\right)^4 \right)^{1/4},
  \qquad
  \nu
  :&= \left( \sum_{n=1}^N \left(\frac{1}{\sigma_n}\right)^4 \right)^{1/4}.
\end{align}
Corollary \ref{corollary:choose-step-to-target-acceptance} shows that if $\sigma_n$ does not decay too fast, one may meet desiderata \ref{desiderata:hmc} (ii) using a step size
\begin{align}
  \label{align:step-size-recommendation}
  \bar h &\approx \frac{1}{\nu} \ 2^{7/4}\sqrt{\Phi^{-1}\left( 1 - \frac{\bar a}{2} \right)},
\end{align}
where $\Phi$ is the normal cumulative distribution function.
The number of leapfrog steps is then proportional to $\sigma_1 / h \propto \sigma_1 \nu = \W$.  Thus $\W$ is a measure of work in a tuned HMC setup.

\begin{figure}
  \centering
  \includegraphics[width=0.98\textwidth]{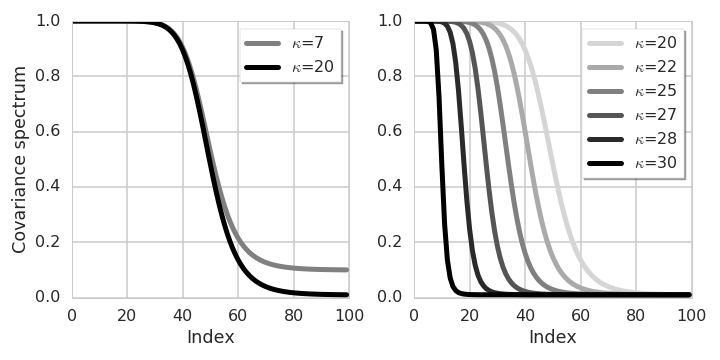}
  \caption{\textbf{Spectra and $\W$.}
  Spectra were generated using $f$ from \eqref{align:decay-func}, and $\W$ was computed.
  \textbf{Left:} Holding the maximal eigenvalue ($\sigma_1$) at one, smaller tails of the spectrum increase $\W$.
  \textbf{Right:} Holding $\sigma_1/\sigma_N$ constant, the worst case is to have one large eigenvalue, and many small ones.}
  \label{fig:many-spectra}
\end{figure}

Since $\W$ involves a ratio of eigenvalues, it is the shape (as opposed to overall scale) of the spectrum that determines the conditioning.
As with the spectral condition number, $\sigma_1/\sigma_N$, $\W$ is minimal when the spectrum is flat, i.e., $\sigma_1^2=\cdots=\sigma_N^2$.
Unlike the spectral condition number, $\W$ is worst when there are many small eigenvalues and at least one large one (see figure \ref{fig:many-spectra}).

\subsection{Numerical estimation of $\W$ and demonstration of main results.}
\label{section:estimation-and-verification-of-kappa}
\begin{figure}
  \centering
  \includegraphics[width=0.49\textwidth]{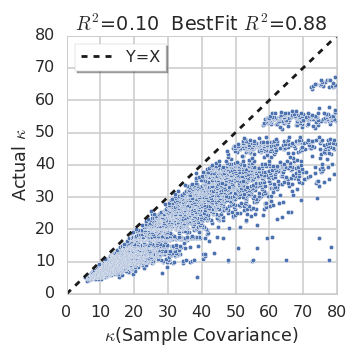}
  \includegraphics[width=0.49\textwidth]{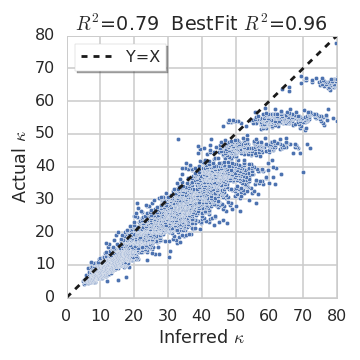} \\
  \includegraphics[width=0.49\textwidth]{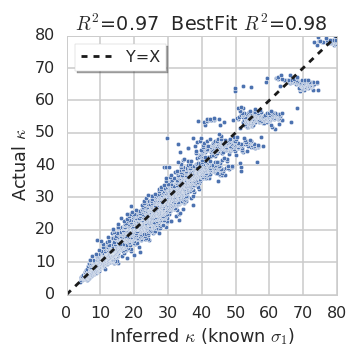}
  \includegraphics[width=0.49\textwidth]{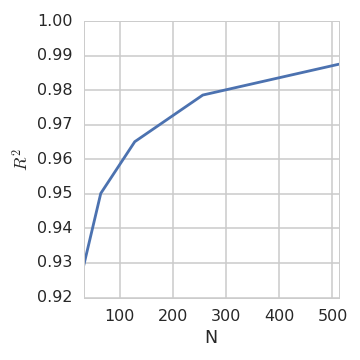}
  \caption{\textbf{Estimation of $\W$ and validation of main result.}
  Random spectra were generated using $f(\calY, m, M, \beta, c)$ from \eqref{align:decay-func}, and the relationship between estimated and actual $\W$ is plotted.  The coefficient of determination $R^2$ of estimated vs.\ actual is shown, as is $R^2$ of the best fit line.
  \textbf{Top Left:} Estimating $\W$ directly from the sample covariance matrix resulted in over-estimation.
  \textbf{Top Right:} Using \eqref{align:step-size-recommendation-rewrite} with estimated $\hat \sigma_1$ gives a fairly accurate estimate of $\W$.  It tends to over-estimate, due to over-estimation of $\sigma_1$.
  \textbf{Bottom Left:} When $\sigma_1$ is known, $R^2\approx1$.  This validates theorem \ref{theorem:integration-error} in the finite $N$ case.
  \textbf{Bottom Right:} $R^2$ vs.\ $N$ when $\sigma_1$ is known.  $R^2$ appears set to converge to $1$, validating theorem \ref{theorem:integration-error}.
  }
  \label{fig:verify}
\end{figure}

A straightforward estimate of $\W$ may be obtained by plugging the sample covariance into \eqref{align:WN}.
  We did this, and it led to a very inaccurate estimate (see figure \ref{fig:verify} ``Sample $\W$'').
The reason being, accurate estimation of a covariance matrix requires many samples (see e.g., discussion of the Wishart case in section \ref{section:how-many-samples}).

We obtained a more accurate estimate by re-writing \eqref{align:step-size-recommendation} as
\begin{align}
  \label{align:step-size-recommendation-rewrite}
  \W :&= \sigma_1 \nu
  \approx \frac{\sigma_1}{\bar h} \, 2^{7/4}\sqrt{\Phi^{-1}\left( 1 - \frac{\bar a}{2} \right)}.
\end{align}
Thus, $\W$ can be estimated by drawing samples with step size $h$, observing the acceptance probability $\hat a$, estimating $\hat\sigma_1\approx\sigma_1$ from the sample covariance, then plugging into \eqref{align:step-size-recommendation-rewrite} (figure \ref{fig:verify} ``Inferred $\W$'').
In experiments, where we know $\sigma_1$ ahead of time, this relation can be used to check theorem \ref{theorem:integration-error} (figure \ref{fig:verify} ``Inferred $\W$ (known $\sigma_1$)'').

To generate random spectra for these numerical tests, we use the set valued function
\begin{align}
  \label{align:decay-func}
  \begin{split}
    f(\calY;\ m, M, c, \beta) :&= 
    \left\{
      \frac{g(y) - \min_{y\in\calY}\{g(y)\}}{\max_{y\in\calY}\{g(y)\} - \min_{y\in\calY}\{g(y)\}} \cdot (M - m) + m
      \st y\in\calY
    \right\},\\
    g(y) :&= 1 / \left( 1 + |y / c|^\beta \right).
  \end{split}
\end{align}
We repeatedly drew $\left\{ \sigma_n \right\}\sim f(\calY;\ m, M, c, \beta)$, for random sets $\calY$ of size\\ $N=32, 64, 128, 256, 512$, each sampled from $\calU(0, 1)$, with {\em minval} $m=1$, {\em maxval} $M\in\left\{ 5, 20 \right\}$,  {\em cutoff} $c\in \left\{ 0.25, 0.75 \right\}$, and {\em power} $\beta\in \left\{ 2, 6 \right\}$.
This means $\sigma_n$ are values of the function $g(y)$ re-scaled to the interval $[m, M]$.
See e.g., figure \ref{fig:many-spectra} for examples.
Since the performance of HMC for Gaussian targets is invariant under isometries (see section 4 of \citep{Neal2012-jd}), it suffices to use these spectra in a Multivariate normal with covariance $C=\mbox{Diag}(\sigma_1^2,\ldots,\sigma_N^2)$.
For each spectrum, we adjust step size $h$ until the acceptance probability is close to either 0.8 or 0.95.  HMC (implemented in \texttt{TensorFlow Probability} \citep{Lao2020-qi}) is then used to draw $S$ samples, with the \emph{oversampling ratio} $S/N\in \left\{ 4, 6, 8, 12, 16, 32 \right\}$.
A total of 4790 random spectra were generated.

\subsection{$\W$ is a condition number on scale matrices}
\label{section:hmc-work:kappa-is-a-cond}
Condition numbers provide worst-case bounds on solutions to linear systems.  For HMC, $\W$ also provides a worst-case result of sorts; the work needed to sample from the most difficult direction corresponding to $\sigma_1$.
Our usage of $\W$ is close to the \emph{stiffness ratio} of a linear ODE, which is just the spectral condition number.  The stiffness ratio can determine the number of time steps needed for convergence to steady state.

Recall the vector norm $\|\cdot\|_2$ and the induced matrix norm (the \emph{spectral norm}):
\begin{align*}
  \|x\|_2 :&= \left( \sum_{n=1}^Nx_n^2 \right)^{1/2},\qquad
  \|A\|_2 := \sup_x \frac{\|Ax\|_2}{\|x\|_2}.
\end{align*}

A condition number quantifies worst case sensitivity of solutions to $Ax = b$ with respect to perturbations of $b$.  For example, consider the perturbed system $A(x + \delta x) = b + \delta b$ for nonsingular $A$.
Elementary steps show that
\begin{align*}
  \frac{\|\delta b\|_2}{\|b\|_2}\leq \|A\|_2\|A^{-1}\|_2 \frac{\|\delta x\|_2}{\|x\|_2}.
\end{align*}
Hence $A\mapsto \|A\|_2\|A^{-1}\|_2$ is a condition number.

A class of matrix norms of interest to us are the Schatten Norms \citep{Horn1990-nf}.  For $r\in [1,\infty]$, the $r^{th}$ Schatten norm of matrix $A$, $\|A\|_{S^r}$, is the vector $r$ norm applied to the singular values of $A$.  For example, with $\left\{ \sigma_1,\ldots,\sigma_N \right\}$ the $N$ singular values of $A\in\Rone^{N\times N}$,
\begin{align*}
  \|A\|_{S^r} :&= \left( \sum_{n=1}^N \sigma_n^r \right)^{1/r}.
\end{align*}
Since $\|A\|_2 = \max_n\left\{ \sigma_n \right\}$, we have $\|A\|_2 \leq \|A\|_{S^r}$.  Therefore
\begin{align}
  \label{align:schatten-condition-bound}
  \frac{\|\delta b\|_2}{\|b\|_2}\leq \|A\|_2\|A^{-1}\|_{S^4} \frac{\|\delta x\|_2}{\|x\|_2},
\end{align}
and $A\mapsto \|A\|_2\|A^{-1}\|_{S^4}$ is a condition number w.r.t. $\|\cdot\|_2$.  As compared with $\|A\|_2\|A^{-1}\|_2$, it is inferior since it provides a looser bound.  The interest for us is that it is equal to $\W$.  Indeed, suppose the covariance matrix is $AA^T$.  Then it's eigenvalues $\sigma_1^2\geq\cdots\geq\sigma_N^2>0$ are by definition the squared singular values of $A$, and
\begin{align}
  \label{align:kappa-condition-number}
  \W :&= \left( \sum_{n=1}^N \left( \frac{\sigma_1}{\sigma_n} \right)^4 \right)^{1/4}
  = \|A\|_2 \|A^{-1}\|_{S^4}
  = \sqrt{\|AA^T\|_2\|(AA^T)^{-1}\|_{S^2}}.
\end{align}
The first equality shows $\kappa$ is a condition number on the \emph{scale matrix} $A$.  The second writes $\kappa$ in terms of the \emph{covariance matrix} $AA^T$, which is often more convenient.

\subsection{Additional properties of $\W$}
By definition, the set of covariance matrices for multivariate normals is the set of symmetric positive definite (SPD) matrices.
Viewing $\W$ as a function of covariance, we have
\begin{align}
  \label{align:kappa-as-a-function-on-GLN}
  \W(C) :&= \sqrt{\|C\|_2 \|C^{-1}\|_{S^2}}.
\end{align}

\begin{lemma}
  \label{lemma:kappa-invariance-properties}
  $\W:SPD\to(0, \infty)$ satisfies
  \begin{enumerate}
    \item[(i)] $\kappa(C)^2 = \lim_{k\to\infty}\sqrt[k]{\Trace{C^k}}\cdot\sqrt{\Trace{C^{-2}}}$
    \item[(ii)] Suppose $A, B$ are non-singular, then $\W(A BB^T A^T) = \W(B^T A^TA B)$
    \item[(iii)] Suppose $U$ is orthogonal and $C$ is SPD, then $\W(U C U^T) = \W(C)$
  \end{enumerate}
\end{lemma}
\begin{proof}
  If $C$ is SPD, its eigenvalues are its singular values $\sigma_1^2\geq\cdots\geq\sigma_N^2>0$, and then $\Trace{C^k} = \sum_{n=1}^N \sigma_n^{2k}$. Taking the limit, we have (i).  To show (ii), use (i) along with the cyclic permutation property of the trace.  (iii) follows from (ii) with $C=BB^T$.
\end{proof}

\subsection{Sequences of spectra}
To study convergence, we must establish a way of taking dimension to infinity.
We draw inspiration from the discretization of a continuous linear operator.  Here, we expect the $N$ point discretization to have singular values $(\sigma_{N1}, \ldots,  \sigma_{NN})$ that are close to singular values of the continuous operator \citep{Hansen1988-ky}.  This arises e.g., in linear inverse problems \citep{Kaipio2006-kd}.  Another example is the discretization of a linear filter in signal processing.
By contrast, \citep{Beskos2013-nd, Betancourt2014-kg, Neal2012-jd} consider a \emph{fixed} set of (possibly non-Gaussian, correlated) random variables, then let $p(x)$ be the law of $N$ \textiid groups of these fixed variables.  This simplification allows them to ignore problems associated with vanishing eigenvalues, but would be unnatural in our setting.

\subsection{Acceptance probabilities for sequences of spectra}
Here we use a random integration time $T_N := \sigma_{N1} T$, where $T\sim\pi$.
Let
\begin{align*}
  \hat\pi(\omega) &:= \int e^{-i\omega t}\pi(t)\dt.
\end{align*}
The bound $|\hat\pi|\leq\hat\pi(0) = 1$ is trivial.  In addition, we impose the regularity condition
\begin{align}
  \label{align:pi-regularity-conditions}
  \begin{split}
    |\hat\pi(\omega)| &\leq C_\pi < 1,
    \qbox{for all}
    |\omega| \geq 2.
  \end{split}
\end{align}
This condition is satisfied e.g., if $\pi$ is a uniform density on any interval.
This ensures each integral appearing in \eqref{align:hN} is uniformly (in $n$) bounded below, and allows derivation of \eqref{align:hN-upper-lower-bound}.

For $\alpha>0$, $N\in\Nat$, define the step sizes
\begin{align}
  \label{align:hN}
  \begin{split}
    h_N :&= \left( 
    \frac{1}{\alpha}\sum_{n=1}^N \frac{1}{(2\sigma_{N,n})^4}\int\sin^2\left( \frac{t}{\sigma_{Nn}} \right)\frac{\pi(t/\sigma_{N1})}{\sigma_{N1}} \dt
    \right)^{-1/4}, \\
    \bar h_N :&= \left( 
    \frac{1}{\alpha}\sum_{n=1}^N \frac{1}{(2\sigma_{N,n})^4} \frac{1}{2}\right)^{-1/4}.
  \end{split}
\end{align}
Note that $\bar h_N$ is, up to a constant, the inverse of $\nu$ from \eqref{align:WN}.

One can show (see section \ref{section:proofs:preliminaries}) that
\begin{align}
  \label{align:hN-upper-lower-bound}
  (1 + C_\pi)^{-1/4}\, \bar h_N
  \leq h_N &\leq
  (1 - C_\pi)^{-1/4} \bar h_N,
\end{align}
so the step sizes differ by at most a constant, and may sometimes be equal (corollary \ref{corollary:simple-step-size}).  Moreover, the proof of theorem \ref{theorem:integration-error} shows the chain is stable as soon as $h_N < 2\sigma_{NN}$.  Then, since (see \eqref{align:step-size-sigma-ratio-uniform-bound} in section \ref{section:proofs:preliminaries}) $h_N / \sigma_{NN}\to0$, the chain is stable for large enough $N$.

We assume the spectra do not decay too rapidly in the sense that
\begin{align}
  \label{align:spectral-decay-assumption}
  \lim_{N\to\infty}
  \sigma_{N1} \left( \sum_{n=1}^N\frac{1}{\sigma_{Nn}^7} \right)
  \left( \sum_{n=1}^N \frac{1}{\sigma_{Nn}^4} \right)^{-3/2}
  = 0.
\end{align}
One can check that \eqref{align:spectral-decay-assumption} holds for any polynomial decrease $\sigma_{N,n}\sim n^{-k}$, but not for exponential $\sigma_{N,n}\sim e^{-n}$.
Note also that \eqref{align:spectral-decay-assumption} provides \emph{uniform} control over the spectra, e.g., it implies $h_N / \sigma_{Nn}\to0$ uniformly in $n$ (see \eqref{align:step-size-sigma-ratio-uniform-bound}).  This allows convergence despite $\left\{\sigma_{Nn}\right\}$ otherwise being unrelated at different $N$.

Our (standard) choice of momentum term, $\|\xi\|^2/2$, means leapfrog integration is invariant under isometry (see section 4 of \citep{Neal2012-jd}). Applying a rotation aligning the axis with eigenvectors of the covariance matrix, the Hamiltonian is diagonalized, and integration error may be studied one component at a time.  Integration error in component $n$, after $\ell$ leapfrog steps is
\begin{align*}
  \delta_{N,n} :&= H(\Psi_{N,n}^\ell) - H(\Psi_{N,n}^0).
\end{align*}
The total integration error is then
\begin{align*}
  \Delta_N :&= \sum_{n=1}^N \delta_{N,n}.
\end{align*}

\begin{theorem}
  \label{theorem:integration-error}
  Given step size $h_N$ from \eqref{align:hN}, integration time $T_N = \sigma_{N1}T$ with $T\sim\pi$ satisfying \eqref{align:pi-regularity-conditions}, and sequences of spectra satisfying \eqref{align:spectral-decay-assumption}, we have convergence in distribution for the HMC (leapfrog) integration error
  \begin{align*}
    \Delta_N &\to \calN\left( \frac{\alpha}{2}, \alpha \right),
  \end{align*}
  for chains in equilibrium.
\end{theorem}
Theorem \ref{theorem:integration-error} is not surprising.  Indeed, since $\Delta_N$ is a sum of $N$ independent random variables, one expects a central limit theorem to hold, provided the scaling given by $h_N$ is correct, and many terms contribute to the sum (as opposed to it being dominated by a few terms).  See e.g.~the CLT for triangular arrays in \citep{Durrett2010-ae}.  The meat of the proof is establishing this scaling (see section \ref{section:proofs:normal-limit}).  Once that is done, assumption \eqref{align:spectral-decay-assumption} ensures many terms contribute to the sum.

The inclusion of the integral $\int\sin^2(\cdot)\dt$ in $h_N$ is ugly.
Unfortunately, it is necessary to handle the case where the spectrum contains significant terms close to $\sigma_{N1}^2$, for which the averaging of $\sin^2(\cdot)$ does not happen.  One simple case where it does is
\begin{corollary}
  \label{corollary:simple-step-size}
  Assume there exists $\delta, C>0$ such that $|\hat \pi(\omega)|\leq C|\omega|^{-\delta}$.  Suppose further $\sigma_{NK}/\sigma_{N1} < r_K$, where $r_K$ is a sequence $\to0$ as $K\to\infty$.
  Then  as $N\to\infty$,
  \begin{align*}
    \frac{\bar h_N}{ h_N}&\to1.
  \end{align*}
\end{corollary}

Corollary \ref{corollary:choose-step-to-target-acceptance} shows the free parameter $\alpha$ may be chosen to achieve desired acceptance rate $\bar a\in (0, 1)$.
\begin{corollary}
  \label{corollary:choose-step-to-target-acceptance}
  Given the hypothesis of theorem \ref{theorem:integration-error}, choose (with $\Phi$ the normal distribution function)
  \begin{align*}
    \alpha :&= 4 \left( \Phi^{-1}\left(1 - \frac{\bar a}{2} \right) \right)^2,
  \end{align*}
  for use in $h_N$.  We then have
  \begin{align*}
    \lim_{N\to\infty} \Exp{a_N(x_j,\xi_j\to \Psi^\ell_{Nn})} = \bar a.
  \end{align*}
  Given hypothesis of corollary \ref{corollary:simple-step-size}, the same result holds with $\bar h_N$ in place of $h_N$.
\end{corollary}
\begin{proof}
  Designate the distributional limit of $\Delta_N$ by $\Delta_\infty\sim\calN(\alpha/2, \alpha)$.
  As in the proof of theorem 3.6 in \citep{Beskos2013-nd}, the boundedness of $u\mapsto 1\wedge e^u$ implies
  \begin{align*}
    \Exp{a_N(x_j,\xi_j\to \Psi^\ell_{Nn})} \to \Exp{1\wedge e^{-\Delta_\infty}}.
  \end{align*}
  This expectation can be found analytically, and is $2\Phi\left( -\sqrt{\alpha} / 2 \right)$.  The result then follows by inverting the relation and applying the continuous mapping theorem \citep{Durrett2010-ae}.
\end{proof}

\section{$\W$ for Large Inverse Wishart Matrices}
\label{section:kappa-for-inv-wishart}
The $\Wishart{N}{S}$ ensemble is that of $N\times N$ random matrices $(1/S)\sum_{s=1}^S (X^s)(X^s)^T$, where each $X^s\sim\calN(0, I_{N})$.  $C\sim\InverseWishart{N}{S}$ if $C^{-1}\sim\Wishart{N}{S}$.  Wishart matrices arise naturally as the $S$-sample covariance matrix of $N$-variate random normals.  Inverse Wishart matrices can result from preconditioning (lemma \ref{lemma:preconditioned-is-inverse-wishart}).

If $N\to\infty$ with the \emph{oversampling ratio} $S / N \to \omega\in(1, \infty)$, then the smallest and largest eigenvalues of a Wishart matrix approach $a := (1 - \omega^{-1/2})^2$ and $b := (1 + \omega^{-1/2})^2$ almost surely \cite{Silverstein1985-hi}. The limiting spectral density is given by the Mar\u{c}enko-Pastur law\citep{Marcenko1967-vo}.
\begin{align}
  \label{align:marcenko-pastur}
  f(x) :&= \left\{
  \begin{matrix}
    \frac{\omega}{2\pi x}\sqrt{(b - x)(x - a)},\quad& a\leq x \leq b,\\
    0 \quad& \mbox{otherwise.}
  \end{matrix}
  \right.
\end{align}
See also figure \ref{fig:inv-wishart-guidelines}.

\begin{figure}
  \centering
  \includegraphics[width=\textwidth]{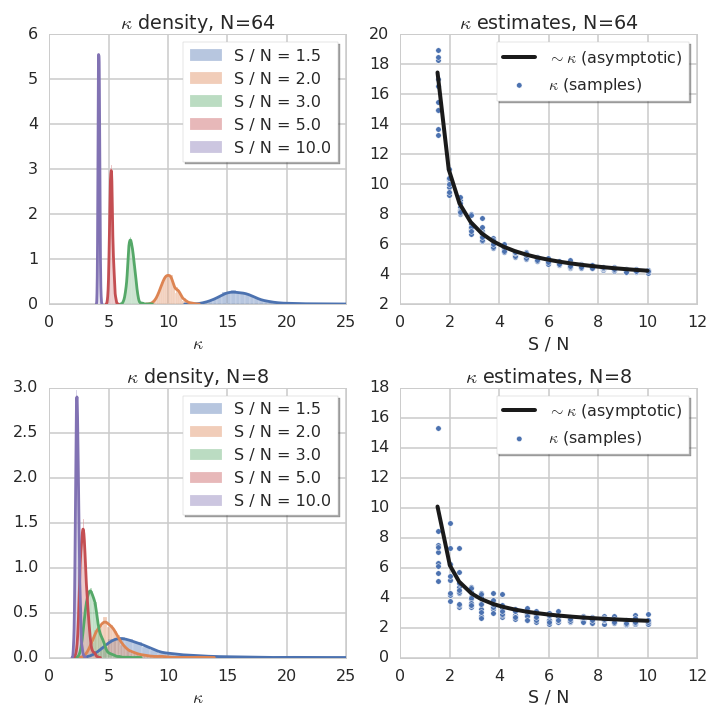}
  \caption{
  \label{fig:kappa-estimates}
  \textbf{Density and asymptotic $\W(C)$, $C\sim\InverseWishart{N}{S}$.}
  \textbf{Top Left:} Density plots when dimension $N=64$.
  Large $S/N$ means $\W$ is closer to $N^{1/4}$, and for $S/N\gtrsim 3$,
  $\W$ concentrates near its mean, so point estimates will be useful.
  \textbf{Top Right:} Estimate vs.\ samples of $\W$ when dimension $N=64$ and $S$ varies.  Here ``asymptotic'' means the large $N$ formula implied by proposition \ref{proposition:inv-wishart-kappa}.  Agreement is very good, except for small $S/N$.
  \textbf{Bottom Left:} When $N=8$, the densities overlap quite a bit, so point estimates will be misleading.
  \textbf{Bottom Right:} When $N=8$, samples of $\W$ vary significantly from the asymptotic $\W$.
  }
\end{figure}

The following proposition gives an asymptotic expression for $\W$ in the inverse Wishart case.  Figure \ref{fig:kappa-estimates} shows good agreement between this expression and sampled $\W$.
\begin{proposition}
  \label{proposition:inv-wishart-kappa}
  If $C\sim\InverseWishart{N}{S}$, and $N\to\infty$ with $S / N\to \omega\in(1, \infty)$, then
  \begin{align*}
    \frac{\W(C)}{N^{1/4}} &\to \frac{(1 + \omega^{-1})^{1/4}}{1 - \omega^{-1/2}}
  \end{align*}
  almost surely.
\end{proposition}
\begin{proof}
  Let $\lambda_1^2\geq\cdots\geq\lambda_N^2>0$ be the eigenvalues of $C^{-1}\sim\Wishart{N}{S}$.  Then,
  \begin{align*}
    \frac{\W(C)^4}{N}
    &= \frac{1}{N}\sum_{n=1}^N \frac{\lambda_N^{-4}}{\lambda_n^{-4}}
    = \lambda_N^{-4}\frac{1}{N}\sum_{n=1}^N\lambda_n^4.
  \end{align*}
  Now, as shown in \citep{Silverstein1985-hi} (main result) and \citep{Bose2008-pl} (remark 5) we have almost sure convergence,
  \begin{align*}
    \lambda_N^2&\to (1 - \omega^{-1/2})^2 \qbox{and}
    \frac{1}{N}\sum_{n=1}^N\lambda_n^4 \to \Exparg{f}{X^2}.
  \end{align*}
  Therefore, almost surely,
  \begin{align*}
    \frac{\W(C)^4}{N}
    &\to \frac{1}{(1 - \omega^{-1/2})^4} \Exparg{f}{X^2}.
  \end{align*}
  The result then follows from
  \begin{align*}
    \Exparg{f}{X^2} &= \int_a^b \frac{\omega}{2\pi x}\sqrt{(b - x)(x - a)} x^2\dx = 1 + \omega^{-1},
  \end{align*}
  which is a straightforward computation involving integrals of the Beta function.
\end{proof}
Convergence of the normalized trace to the moment is a result of remark 5 in \citep{Bose2008-pl}, which gives \as convergence of the empirical spectral density function.  Convergence of this type holds for broad class of matrices \citep{Bai1999-dh}, \citep{Bose2010-wh}.

\section{Preconditioning HMC}
\label{section:preconditioning}
The results of section \ref{section:hmc-work} show a clear sampling advantage for problems where the spectrum is as close to ``flat'' as possible.  Here we consider techniques where a diffeomorphism $F$ transforms the random variable $X$ into $Z := F^{-1}(X)$, with a hope that $Z$ is easier to sample from.
This is not a new idea.  Linear transformations have been considered as far back as \citep{Neal2012-jd}.  Trainable nonlinear mappings seem to have been introduced by \citep{Parno2018-ch}, where variational inference is used to find $F$.  It has since been developed further in \citep{Peherstorfer2019-ro}, which considers mappings based on low-fidelity approximations to the posterior, and \citep{Hoffman2019-pd}, where the preconditioner is a neural network.

To formalize this procedure, let us start with $X\sim p_X(x)$, and a diffeomorphism $F$, which transforms $X\mapsto Z = F^{-1}(X)$.  Equivalently, the density $p_X$ is transformed by the \emph{pushforward}
\begin{align}
  \label{align:pushfwd}
  p_Z(z) &= \pushfwd{F^{-1}}p_X(z) := p_X(F(z))|\det(DF(z))|.
\end{align}
Above, $DF$ is the matrix of partial derivatives, $(DF)_{ij} = \p F_j/\p z_i$.
Using HMC, we sample from the density $p_Z$, producing $Z^1,\ldots,Z^K$.  Transforming back, $X^k := F(Z^k)$, and we have samples from $p_X$ as desired.

In the Gaussian case, $p_X\sim\calN(0, AA^T)$, the linear preconditioner induced by a matrix $F$ transforms the covariance and $\W$ as follows:
\begin{align}
  \label{align:linear-preconditioning-transformations}
  \begin{split}
    AA^T & \mapsto (F^{-1}A)(F^{-1}A)^T = (F^{-1})AA^T(F^{-1})^T \\
    \kappa(AA^T) & \mapsto \|F^{-1}A\|_2\|(F^{-1}A)^{-1}\|_{S^4}.
  \end{split}
\end{align}


\subsection{How Many Samples are Enough?}
\label{section:how-many-samples}
The preconditioning techniques we will consider have an upfront cost:  Either in obtaining some number of preliminary samples (which should be thrown away), or in solving an optimization problem.  It's fair to ask whether the subsequent speed-up is worth it.  That depends on how many final samples are needed.  Here we present two cases where the ratio of samples to dimension, or \emph{oversampling ratio}, $\omega := S / N$, is required to be 10 or more.

First, the estimation of component means and variance is done with summations of the form $(1 / S)\sum_{s=1}^S Y^s$, resulting in relative error on each component of $O(1 / \sqrt{S})$.  Thus, relative error of size $\eps$ requires $S\sim O(\eps^{-2})$.  This is independent of dimension $N$, but for moderate $N\approx100$, and reasonable $\eps\approx 0.025$, we will need $S\approx 1600$, which implies $\omega := S / N \approx 16$ is required.

Second, consider the error in reconstructing the covariance spectrum from the sample covariance matrix $\Chat := (1 / S)\sum_{s=1}^S (X^s)(X^s)^T$.  If $X^s\sim_\iid\calN(0, I_N)$, we have $\Chat \sim\Wishart{S}{N}$, and for the bulk of the spectrum to be within $\eps$ of the correct values (which are all one), we must have $(1 + \omega^{-1/2}) < 1 + \eps$, which implies $\omega > 4 / \eps^2$ (see section \ref{section:kappa-for-inv-wishart}).

\subsection{Preconditioning with Sample Covariance}
\label{section:preconditioning:sample-cov}
Here we consider a choice to be made by a practitioner who has gathered $S \geq N$ burn-in samples $(X^1,\ldots,X^S)$.  They could continue gathering samples until a goal of $S_f > S$ ``final'' are obtained.  Alternatively, they could form the sample covariance $\Chat:= (1/S)\sum_{s=1}^S (X^s)(X^s)^T$, precondition with its Cholesky factor $\Lhat$, throw away the first $S$ samples, then gather $S_f$ final samples.

Abusing notation, we let $\W_0$ be the initial condition number, and $\W(S)$ be the condition number after preconditioning with $S$ samples.  Assuming the sampling rate is proportional to $1 / \W$, then the total time $\tau$ to obtain $S_f$ samples is
\begin{align*}
  \tau &\propto S \W_0 + S_f \W(S),
\end{align*}
which has extremal points $\d \tau/\d S = 0$ at $S^\ast$ whenever
\begin{align}
  \label{align:optimal-schedule-condition}
  \frac{d\W}{\d S}(S^\ast) &= -\frac{\W_0}{S_f}.
\end{align}

This suggests continuing to draw burn-in samples until $\d\W/\d S \leq - \W_0 /S_f$, at which time updates may stop and $S_f$ samples can be drawn.  The speedup is
\begin{align}
  \label{align:optimal-schedule-speedup}
  \frac{S_f\W_0}{S\W_0 + S_f\W(S)}.
\end{align}

Assuming the burn-in samples are i.i.d., we will obtain an approximate expression for the value of $S / N$ at which \eqref{align:optimal-schedule-condition} is met.
First however, we must establish
\begin{lemma}
  \label{lemma:preconditioned-is-inverse-wishart}
  Suppose $(X^1,\ldots,X^S)$ are i.i.d., and HMC sampling of $X\sim\calN(0, C)$ is preconditioned with the $S$-sample Cholesky factor $\Lhat$.  Then the preconditioned $\W$ follows the law of $\W(C)$, for $C\sim\InverseWishart{S}{N}$.
\end{lemma}
\begin{proof}
  The preconditioned covariance is $\Lhat^{-1}C\Lhat^{-T}$.  Due to lemma \ref{lemma:kappa-invariance-properties}, $\W(\Lhat^{-1}C \Lhat^{-T}) = \W(L^T\Chat^{-1}L)$.  Since
  \begin{align*}
    L^T\Chat^{-1}L
    &= \left( \frac{1}{S}\sum_{s=1}^S (L^{-1}X^s)(L^{-1}X^s)^T \right)^{-1},
  \end{align*}
  and $L^{-1}X^s\sim\calN(0, I_N)$, we see that $L^T\Chat^{-1}L\sim\InverseWishart{S}{N}$, which completes the proof.
\end{proof}

To check \eqref{align:optimal-schedule-condition} we will use
lemma \ref{lemma:preconditioned-is-inverse-wishart}, and proposition \ref{proposition:inv-wishart-kappa}.
That is, we start with the approximation
\begin{align}
  \label{align:asymptotic-kappa-s}
  \kappa(S)
  & \approx g_N(S) := N^{1/4}\ \frac{\left(1 + \frac{N}{S}\right)^{1/4}}{1 - \sqrt{\frac{N}{S}}}.
\end{align}
Then, \eqref{align:optimal-schedule-condition} is approximately satisfied when $\W'(S) \approx g_N'(S) = -\W_0/S_f$, which is equivalent to finding $S/N$ such that
\begin{align}
  \label{align:optimal-schedule-approximation}
  U\left( \frac{S}{N} \right) &= \frac{N^{1/4}}{\W_0} \ \frac{S_f}{N},
  \qbox{where}
  U(\omega) := \frac{4(\omega^{1/2} - 1)^2 (\omega^2 + \omega)^{3/4}}{2\omega + \omega^{1/2} + 1}.
\end{align}
This expresses an optimality condition on the burn-in oversampling ratio $S/N$ in terms of the final oversampling ratio $S_f/N$ and the rescaled initial condition number $\W_0/N^{1/4}$.

\begin{figure}
  \centering
  \includegraphics[width=\textwidth]{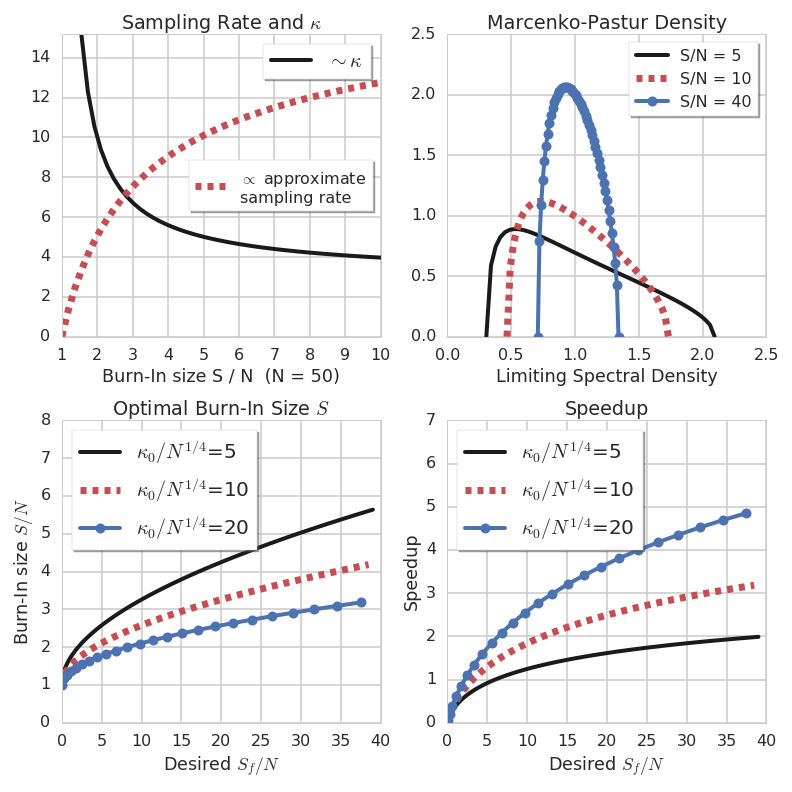}
  \caption{
  \textbf{Sample Covariance Preconditioning:} Charts to help decide how many samples $S$ to use for burn-in.
  {\bf Top Left:} Approximation of $\kappa(S)$, from proposition \ref{proposition:inv-wishart-kappa}.  As always, $\W$ is (approximately) inversely proportional to the sampling rate for HMC.  We conclude larger burn-in size $S$ leads to faster sampling rate.
  {\bf Top Right:} Limiting spectral density for Wishart ensemble, \eqref{align:marcenko-pastur}, for three different $S/N$ values.  The spectrum has a fair bit of spread, even with 40x oversampling.
  {\bf Bottom Left:} The optimal burn-in oversampling ratio $S/N$, as a function of the final desired oversampling ratio $S_f/N$ for different values of $\W_0/N^{1/4}$.  This is the graph $\left\{ (U(\omega)\W_0/N^{1/4}, \omega) \st \omega\in(0, 5) \right\}$, with $U(\omega)$ defined in \eqref{align:optimal-schedule-approximation}.
  {\bf Bottom Right:} Speedup, as defined by \eqref{align:optimal-schedule-speedup}, with $\W(S)$ approximated by \eqref{align:asymptotic-kappa-s}.
    \label{fig:inv-wishart-guidelines}
  }
\end{figure}

These steps are put together in figure \ref{fig:inv-wishart-guidelines}.  For example, suppose $N=50$, $\W_0/N^{1/4}=10$, and uncertainty in sample covariance needs to be less than 25\%.  Then, the ``Marcenko-Pastur Density'' plot shows $S_f/N\approx40$ is required, the ``Optimal Burn-In Size $S$'' plot shows $S/N\approx4$ should be used, and finally, the ``Speedup'' plot shows our expected speedup is 3.

\begin{figure}
  \centering
  \includegraphics[width=0.32\textwidth]{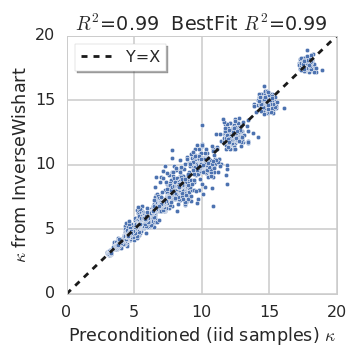}
  \includegraphics[width=0.32\textwidth]{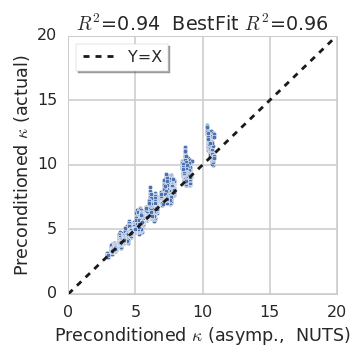}
  \includegraphics[width=0.32\textwidth]{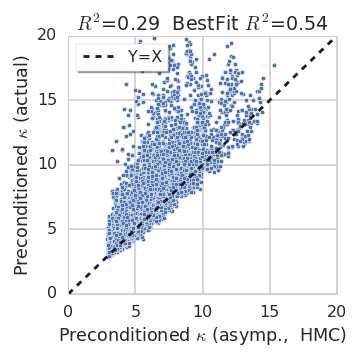}
  \caption{\textbf{Sample Covariance Preconditioning : Theory vs. Numerics.}
  Random spectra were generated using $f(\calY; m, M, \beta, c)$ from \eqref{align:decay-func}, and the relationship between theoretical and actual $\W$ is plotted.  The coefficient of determination $R^2$ of estimated vs.\ actual is shown, as is $R^2$ of the best fit line.
  {\bf Left:} $\W$ after preconditioning with $S$ \textiid samples in dimension $N$ (various $S$ and $N$) is close to $\W$ obtained via samples from $\InverseWishart{N}{S}$, as lemma \ref{lemma:preconditioned-is-inverse-wishart} dictates.  The scatter plots of course don't match perfectly, since samples are random.
  {\bf Center:} The function $g_N(S)$ (with $S$ equal to the mean effective sample size) from \eqref{align:asymptotic-kappa-s}, which approximates $\W$, is compared with the actual $\W$ obtained by preconditioning with $S$ effective samples from a NUTS chain.  Agreement is good.
  {\bf Right:} Same as Center, but using an HMC chain \emph{without} NUTS.  Agreement is not so good.
  }
  \label{fig:wishart-verify}
\end{figure}

To enact these steps in practice, one needs both an estimate of $\W_0$, and $S$ \textiid samples.  The estimate of $\W_0$ can be obtained using the steps in section \ref{section:estimation-and-verification-of-kappa}.  However, \textiid samples are presumably impossible to come by, else we would just use them.  As a supplement, we suggest gathering $S$ HMC samples, then using the effective sample size (ESS) in place of $S$ in \eqref{align:optimal-schedule-approximation} \citep{Robert2004-so}.  Our experiments found ESS to be a good substitute, but \emph{only} if the original samples were obtained using the No-U-Turn Sampler (NUTS) \citep{Hoffman2014-vt}.  See figure \ref{fig:wishart-verify}.  This is unfortunate, since the nice estimate of $\W_0$ (as per section \ref{section:estimation-and-verification-of-kappa}) falls apart with NUTS, due to its non-trivial acceptance criteria.
Our suggested remedy is to obtain a small number of additional non-NUTS samples, and use these to estimate $\W_0$.  This is possible, since estimation of $\W_0$ only requires an estimate of the acceptance probability.

\subsection{Preconditioning by way of Variational Inference}
\label{section:preconditioning:vi}
In variational inference, parameters $\theta$ are tuned to minimize a loss function involving a parameterized distribution $q(\cdot;\theta)$ and the target $p$.  For example, the \emph{reverse KL divergence} is
\begin{align}
  \label{align:reverse-kl}
  \KL{q}{p} &= \int\log\left[ \frac{q(x;\theta)}{p(x)} \right]q(x;\theta)\dx.
\end{align}
We will always choose $q$ to be a pushforward (see \eqref{align:pushfwd}) of the standard normal.  That is, $q=\pushfwd{F}\phi$, for $\phi\sim\calN(0, I_N)$.
It follows that
\begin{align}
  \label{align:kl-under-diffeomorphism}
  \KL{q}{p}
  &= \KL{\pushfwd{F}\phi}{p}
  = \KL{\phi}{(\pushfwd{F})^{-1}(p)},
\end{align}
which leads to the approximation
\begin{align}
  \label{align:reverse-kl-summation}
  \KL{q}{p}
  &\approx \frac{1}{K}\sum_{k=1}^K \log\left[ \frac{\phi(z^k)}{p(F(z^k))|DF(z)|} \right]
  = \frac{1}{K}\sum_{k=1}^K \log\left[ \frac{q(F(z^k))}{p(F(z^k))} \right]
  ,\qquad z^k \sim\phi.
\end{align}
If $F$ is smooth and $\theta\mapsto \KL{q}{p}$ is convex, \eqref{align:reverse-kl} may be minimized by stochastic gradient descent using \eqref{align:reverse-kl-summation}. This is an example of \emph{sample path optimization} \citep{Amaran2016-ze}.  Since the summands are log probabilities, \eqref{align:reverse-kl-summation} is usually found to be stable.

The ``reverse'' moniker is attached to \eqref{align:reverse-kl} to differentiate it from \emph{forward KL divergence},
\begin{align}
  \label{align:forward-kl}
  \KL{p}{q} &= \int\log\left[ \frac{p(x)}{q(x;\theta)} \right]p(x)\dx.
\end{align}
Since presumably $p(x)$ is \emph{not} easy to sample from, a stable ``log space'' formula analogous to \eqref{align:reverse-kl-summation} cannot be used to approximate \eqref{align:forward-kl}.

Equation \eqref{align:kl-under-diffeomorphism} shows that, if our minimization results in small KL divergence, then $(\pushfwd{F})^{-1}p$ is close to the well-conditioned unit Gaussian, in the sense of KL divergence.
Unfortunately, this does not imply $(\pushfwd{F})^{-1}p$ has well-conditioned covariance.  Indeed, if $p$ is Gaussian, then, with $\lambda_n^2$ the eigenvalues of the preconditioned covariance $WW^T$, one can check that, up to additive and multiplicative constants,
\begin{align}
  \label{align:kl-in-terms-of-eigenvalues}
  \begin{split}
    \KL{p}{q} &\propto \|W\|_F^2 - \log|\det(WW^T)| 
    = \sum_{n=1}^N \left( \lambda_n^2 - \log(\lambda_n^2) \right), \\
    \KL{q}{p} &\propto \|W^{-1}\|_F^2 - \log|\det( (WW^T)^{-1})| 
    = \sum_{n=1}^N \left( \lambda_n^{-2} - \log(\lambda_n^{-2}) \right).
  \end{split}
\end{align}
Clearly, minimizing forward or reverse KL is different than minimizing $\W$.

\subsubsection{Diagonal preconditioning}
\label{section:preconditioning:vi:diagonal}
Here we consider preconditioning a Gaussian $\calN(0, C)$ with a diagonal matrix $D$.  The covariance and $\W$ transform as in \eqref{align:linear-preconditioning-transformations}.

Minimizing reverse KL over the set of diagonal matrices (see \eqref{align:kl-in-terms-of-eigenvalues}) gives us
\begin{align}
  \label{align:reverse-kl-diagonal-minimizer}
  D_{ii}^2 = 1 / (C^{-1})_{ii},
\end{align}
while minimizing forward KL gives us
\begin{align}
  \label{align:forward-kl-diagonal-minimizer}
  D_{ii}^2 = C_{ii},
\end{align}
which is just a diagonal matrix with the component-wise variances.  In this case, preconditioning is equivalent to re-scaling the axis so that $X$ has unit standard deviation.

As diagonal preconditioners, both forward and reverse KL exhibit a scale invariance,
the proof of which follows directly from \eqref{align:reverse-kl-diagonal-minimizer} and \eqref{align:forward-kl-diagonal-minimizer}.
\begin{lemma}
  \label{lemma:KL-scale-invariance}
  Suppose forward/reverse KL preconditioning of $X\sim\calN(0, C)$ results in preconditioned covariance $\Omega$.  Then, for any positive diagonal matrix $\tilde D$, forward/reverse KL diagonal preconditioning of $X\sim\calN(0, \tilde D C \tilde D)$ also leads to preconditioned covariance $\Omega$.
\end{lemma}

Consider the following choices:
\begin{enumerate}
  \item[(i)] Do not precondition, and sample directly from the target $p(x)$.
  \item[(ii)] Precondition with $D^{-1}$, where $D$ is obtained by minimizing $\KL{q}{p}$, (i.e., reverse KL), for $q(x)\sim\calN(\mu, D^2)$.
  \item[(iii)] Precondition with $D^{-1}$, where $D$ is obtained by minimizing $\KL{p}{q}$, (i.e., forward KL), for $q(x)\sim\calN(\mu, D^2)$.
\end{enumerate}
In the proceeding sections, we will show realistic scenarios where each method is better than the other two.  Before proceeding, we point out some practical considerations.  Since forward KL is often unstable and cannot be minimized directly, (iii) is done by estimation of the component-wise standard deviation.  If this must be done by sampling, then it somewhat defeats the purpose of preconditioning.  Regarding (ii), setting up a variational problem is not too hard once the target $p(x)$ is built, and software packages exist to make this easier \citep{Dillon2017-zq}.  This does however incur a one time development cost that may be too great for the problem at hand.

\begin{figure}
  \centering
  \includegraphics[width=0.98\textwidth]{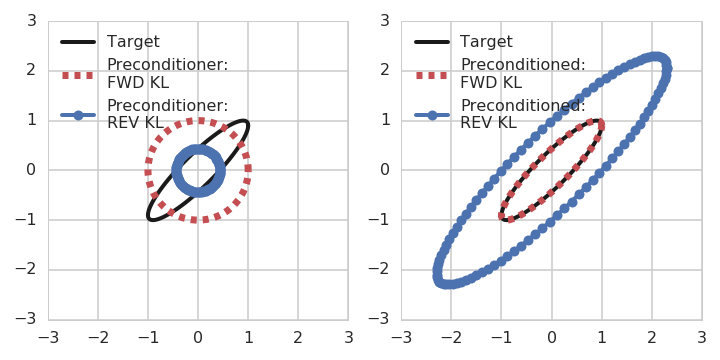}
  \caption{\textbf{Forward and reverse KL, preconditioner and preconditioned.}
    The target $p(x)\sim\calN(0, C_n)$, with $C_n$ from \eqref{align:Gamma-n} and ($\rho=0.9$).  The variational distribution $q(z)\sim\calN(0, D_n^2)$, where $D_n$ is diagonal. Both forward and reverse KL were minimized to find $D_n$.
    {\bf Left:} The one standard deviation iso-line of $q(z)$ is plotted.  Both are circular due to symmetry, but forward KL chooses a larger sphere (of radius 1).
    {\bf Right:} Iso-lines after preconditioning by $D_n$ (which yields $\sim\calN(0, D_n^{-1}C_n D_n^{-1})$).  Reverse KL results in a much larger preconditioned spectrum.}
  \label{fig:fwd-rev-ellispses}
\end{figure}

\subsubsection{Diagonal preconditioning of correlated diagonal blocks}
\label{section:preconditioning:vi:correlated-blocks}
A simple covariance comprised of 2x2 blocks provides a demonstration of cases where forward KL preconditions better than reverse KL, which, depending on the blocks, performs better or worse than doing nothing.
We also see that neither forward nor reverse KL is optimal.

\begin{figure}
  \centering
  \includegraphics[width=0.98\textwidth]{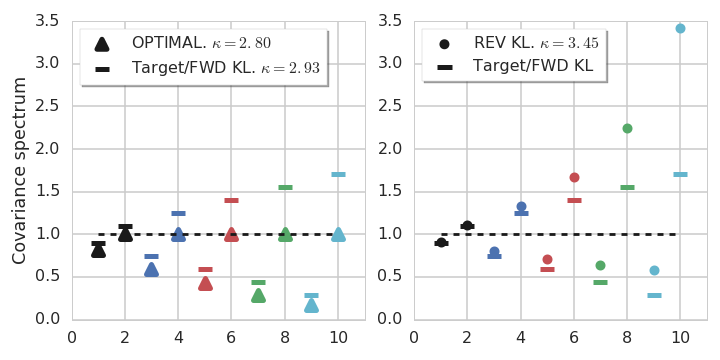}
  \caption{
  \textbf{Preconditioned spectra in the case of 5 correlated 2x2 blocks.}
  Preconditioner choices (OPTIMAL vs. REV KL.) are plotted against the target, which is the same as FWD KL preconditioning.  The spectrum comes in pairs of eigenvalues, which are scaled together by the preconditioners.
  {\bf Left:} The optimal preconditioner pushes the larger member of each pair to 1.
  {\bf Right:} Reverse KL (almost) pushes the smaller of each pair to 1.}
  \label{fig:correlated-block-spectra}
\end{figure}

For $\rho_n\in(0,1)$, let covariance be given by the block-diagonal matrix
\begin{align}
  \label{align:Gamma-n}
  C &= C_1\oplus C_2\oplus\cdots\oplus C_N,\quad
  C_n := \left( 
  \begin{matrix}
    1 & \rho_n \\
    \rho_n & 1
  \end{matrix}
  \right),
\end{align}
which has eigenvalues $\left\{1 \pm \rho_n\st n=1,\ldots,N\right\}$.  By symmetry, the optimal preconditioner $D$, for forward or reverse KL, will be partitioned as
\begin{align*}
  D &= D_1\oplus\cdots\oplus D_N,
  \quad
  D_n := \left( 
  \begin{matrix}
    d_n & 0 \\
    0 & d_n
  \end{matrix}
  \right).
\end{align*}
This leads to the preconditioned covariance
\begin{align*}
  D^{-1}C D^{-1} &= d_1^{-2}C_1\oplus\cdots\oplus d_N^{-2}C_N,
\end{align*}
with spectrum
\begin{align*}
  \Lambda := \left\{
  \frac{1 + \rho_1}{d_1^2}, \frac{1 - \rho_1}{d_1^2},
  \cdots,
  \frac{1 + \rho_N}{d_N^2}, \frac{1 - \rho_N}{d_N^2},
  \right\}.
\end{align*}
Thus, each pair of eigenvalues, $\left\{ 1\pm\rho_n \right\}$ is moved up and down together by the preconditioner.  As seen below and in in figure \ref{fig:correlated-block-spectra}, the optimal preconditioner moves the larger of the two from every block to the same level.  Reverse KL to some extent does the opposite, moving the smaller of each pair to a similar level.  This is expected, since as illustrated in figure \ref{fig:fwd-rev-ellispses}, the scale of the reverse KL variational solution is mostly determined by the smallest scale of the target.

In the forward KL, reverse KL, and the optimal choices of $d_n$ that follow, $(1 + \rho_1) / d_1^2\geq (1 \pm \rho_n)/d_n^2$, for $n=2,\ldots,N$.  Therefore, all three choices will have
\begin{align}
  \label{align:W1-correlated-blocks}
  \begin{split}
    \W(\Lambda)^4 &= 
    \sum_{n=1}^N \left( \frac{d_n^2}{d_1^2} \right)^2
    \left[ 
    \left( \frac{1 + \rho_1}{1 + \rho_n} \right)^2 + \left( \frac{1 + \rho_1}{1 - \rho_n} \right)^2
    \right].
  \end{split}
\end{align}
Referring to \eqref{align:reverse-kl-diagonal-minimizer}, \eqref{align:forward-kl-diagonal-minimizer}, the minimizing $d_n^2$ for forward KL will be identically $1$, and for reverse KL will be $1 - \rho_n^2$.  Thus
\begin{align*}
  \W(\Lambda_{rev})^4 &= 
    \sum_{n=1}^N \left( \frac{1 - \rho_n^2}{1 - \rho_1^2} \right)^2
    \left[ 
    \left( \frac{1 + \rho_1}{1 + \rho_n} \right)^2 + \left( \frac{1 + \rho_1}{1 - \rho_n} \right)^2
    \right] \\
  \W(\Lambda_{fwd})^4 &= 
    \sum_{n=1}^N 
    \left[ 
    \left( \frac{1 + \rho_1}{1 + \rho_n} \right)^2 + \left( \frac{1 + \rho_1}{1 - \rho_n} \right)^2
    \right],
\end{align*}
and therefore $\W(\Lambda_{fwd})\leq \W(\Lambda_{rev})$.

Minimizing $\W$ over all $d_n$ we find $d_n^2\propto 1 + \rho_n$, so that
\begin{align*}
  \W(\Lambda_{OPT})^4 &= 
    \sum_{n=1}^N
    \left[ 
    1 + \left( \frac{1 + \rho_n}{1 - \rho_n} \right)^2
    \right].
\end{align*}

Since reverse KL is a practical method, it is disappointing to see that doing nothing (forward KL had $D=I$) performs better.  In practice however, the situation is often closer to 
\begin{align*}
  C_n :&= \gamma_n^2 \left( 
  \begin{matrix}
    1 & \rho_n \\
    \rho_n & 1
  \end{matrix}
  \right),
\end{align*}
for $\gamma_1^2\geq\cdots\geq\gamma_N^2>0$.  The results for forward and reverse KL will be the same as if $\gamma_n\equiv1$ due to the scale invariance lemma \ref{lemma:KL-scale-invariance}, but ``doing nothing'' yields a baseline of (with $\beta := \max_n\left\{ \gamma_n^2(1 + \rho_n) \right\}$)
\begin{align*}
  \W^4 :&=
  \left[ 
  \frac{\beta^2}{\gamma_n^4(1 + \rho_n)^2} +
  \frac{\beta^2}{\gamma_n^4(1 - \rho_n)^2}
  \right]
  \geq
  \sum_{n=1}^N 
  \frac{\gamma_1^4}{\gamma_n^4}
  \left[ 
  \left( \frac{1 + \rho_1}{1 + \rho_n} \right)^2 +
  \left( \frac{1 + \rho_1}{1 - \rho_n} \right)^2
  \right].
\end{align*}
So if $\gamma_1\gg\gamma_n$ is large enough, preconditioning with reverse KL \emph{does} improve upon doing nothing.  Forward KL would still be superior.

\subsubsection{Diagonal preconditioning of random matrices}
Here we compare preconditioner options (Forward or Reverse KL, or ``Do Nothing'') applied to 100x100 random matrices of different types.  Each option is best some fraction of the time.

The \emph{Wishart} random matrices are constructed by (i) letting $A\in\Rone^{100\times 200}$ be composed of i.i.d.\ unit normal entries, then (ii) setting $C := AA^T$.  The \emph{inverse Wishart} matrices are constructed by inverting a Wishart matrix.  The \emph{rotated scale} matrices are constructed by (i) making the scale matrix $\Lambda := \mbox{Diag}(\sigma_1^2,\ldots,\sigma_{100}^2)$, where $\{\sigma_1,\ldots,\sigma_{100}\} = f(\{1,\ldots,100\}, m=1, M=5, \beta=4, c=\texttt{cutoff})$ from \eqref{align:decay-func}, then (ii) rotating with a random orthogonal matrix generated with \texttt{scipy.stats.ortho\_group}, i.e.~$C:= U\Lambda U^T$ \citep{Virtanen2019-pu}.  Three varieties of rotated scale matrix were constructed by choosing $\texttt{cutoff}=0.05, 0.1, 0.2$, which means the scales have approximately 5\%, 10\%, or 20\% of the values near the maximum (of 5), and the remainder near the minimum (of 1).  For each of the 5 matrix types, 1000 different random matrices were generated, and the fraction of the time each preconditioner type leads to lower $\W$ is shown in table \ref{table:preconditioner-winner-comparison}.

The Wishart/inverse Wishart matrices have a fairly regular structure, and forward KL is the ``winner'' more and more often as $N$ increases. The results for rotated scale matrices varied considerably as the dimension or the parameters $(M,\beta)$ were changed.  We warn the reader \emph{not} to draw broad conclusions or trends from table \ref{table:preconditioner-winner-comparison}.
\begin{table}[h!]
  \centering
\begin{tabular}{|l||c|c|c|c|c|}
 \hline 
            & Wishart & InvWishart & RS (5\%) & RS (10\%) & RS (20\%) \\
 \hline\hline 
 Do Nothing & 3\% & 0\% & 0\% & 20\%  &100\% \\
 \hline 
 Fwd KL & 91\% & 100\% & 88\% & 1\%  &0\% \\
 \hline 
 Rev KL & 6\% & 0\% & 12\% & 78\%  &0\% \\
 \hline 
\end{tabular}
\caption{For Wishart, inverse Wishart, and rotated scale (RS) random 100x100 matrices, the percentage of the time each preconditioning method had the lowest $\W$ is tabulated.  The magnitude of the differences was small, about 10\% at most.\label{table:preconditioner-winner-comparison}}
\end{table}

\subsubsection{Diagonal plus low-rank preconditioning}
\begin{figure}
  \centering
  \includegraphics[width=0.49\textwidth]{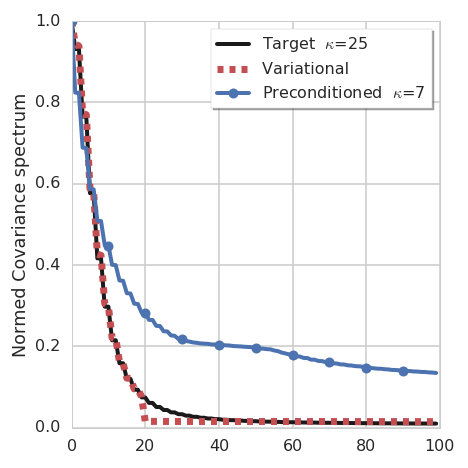}
  \includegraphics[width=0.49\textwidth]{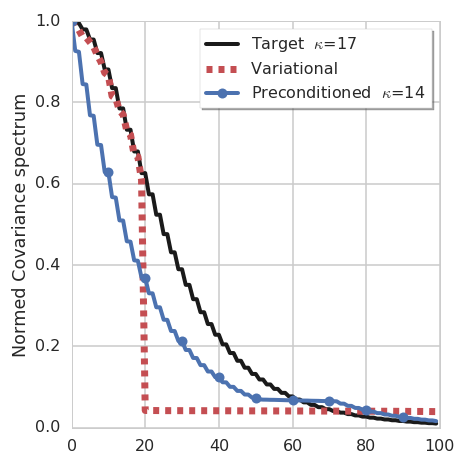}
  \caption{\textbf{Preconditioning with a low-rank update.}  Normalized covariance spectra of the Target $p\sim\calN(0, LL^T)$, variational model $q\sim\calN(0, D + UU^T)$, and the preconditioned target.
  {\bf Left:} $LL^T$ had 10 large eigenvalues, and $U$ was a rank 20 update.  The trained preconditioner $D + UU^T$ had about 10 large eigenvalues, which were used to reduce the largest eigenvalues in the preconditioned matrix.  
  {\bf Right:} Here $LL^T$ had about 40 large eigenvalues, and the rank 20 update could not reduce the size of them all.}
  \label{fig:low-rank}
\end{figure}

As demonstrated in figure \ref{fig:many-spectra}, having a few large eigenvalues and many small ones is especially bad for $\W$.  To mitigate these situations, we consider a low-rank update to a diagonal preconditioner.  Specifically, we choose variational distribution $q\sim\calN(0, FF^T)$, where $F = D + UU^T$, $D\in\Rone^{N\times N}$ is diagonal, and $U\in\Rone^{N\times K}$.  Both $D$ and $U$ were trained to minimize $\KL{q}{p}$, where $p\sim\calN(0, LL^T)$ and $L$ is circulant.  This provides some correlation that cannot be matched with a diagonal preconditioner.  The spectrum of $L$ was chosen using \eqref{align:decay-func} so that it was a low pass filter with some cutoff.  As expected, when the rank of $U$ was larger than the cutoff, preconditioning worked.  When the rank of $U$ was less than the cutoff, large eigenvalues remained and $\W$ was barely reduced by preconditioning.  See figure \ref{fig:low-rank}.

A word of caution: We also found that when the circulant matrix $L$ had very small eigenvalues, the corresponding extreme correlations in $X\sim p(x)$ were too much for the optimization procedure to handle, and instabilities arose.

\section{Proof of Convergence Results}
\label{section:proofs}
In this section we prove theorem \ref{theorem:integration-error} and corollary \ref{corollary:simple-step-size}.

Denote by $\pi_N$ the density of $T_N := \sigma_{N1}T$ (recall $T\sim\pi$).  That is,
\begin{align*}
  \pi_N(t) :&= \pi\left( \frac{t}{\sigma_{N1}} \right)\cdot \frac{1}{\sigma_{N1}}.
\end{align*}

\subsection{Two relations involving step size and scales}
\label{section:proofs:preliminaries}

First, we show \eqref{align:hN-upper-lower-bound}.  To that end, the distinction between $h_N$ and $\bar h_N$ is the replacement of $\int\sin^2(\cdot)\dt$ by its average, $1/2$.  Let
\begin{align*}
  R_{Nn} :&= \frac{1}{2} - \int \sin^2\left( \frac{t}{\sigma_{Nn}} \right)\pi_N(t)\dt
  = \frac{1}{4}\left[ \hat\pi\left( 2\frac{\sigma_{N1}}{\sigma_{Nn}} \right) + \hat \pi \left( -2\frac{\sigma_{N1}}{\sigma_{Nn}} \right) \right],
\end{align*}
which, due to \eqref{align:pi-regularity-conditions}, satisfies
\begin{align}
  \label{align:sin-integral-1}
  \begin{split}
    |R_{Nn}| &\leq \frac{C_\pi}{2} < \frac{1}{2}.
  \end{split}
\end{align}
This means
\begin{align*}
  \left| \bar h_N^{-4} - h_N^{-4}  \right| &\leq C_\pi \bar h_N^{-4},
\end{align*}
and therefore
\begin{align*}
  (1 + C_\pi)^{-1/4}\, \bar h_N
  \leq h_N &\leq
  (1 - C_\pi)^{-1/4} \bar h_N,
\end{align*}
which is exactly \eqref{align:hN-upper-lower-bound}.

Second, we show the uniform limit
\begin{align}
  \label{align:step-size-sigma-ratio-uniform-bound}
  \lim_{N\to\infty}
  \frac{h_N}{\sigma_{Nn}}
  \leq
  \lim_{N\to\infty}
  \frac{h_N}{\sigma_{NN}} = 0.
\end{align}
To that end, since $\sigma_{NN}\leq \sigma_{Nn}$, we have
\begin{align*}
  \left( \frac{h_N}{\sigma_{Nn}} \right)^6
  \leq\left( \frac{h_N}{\sigma_{NN}} \right)^6
  &\leq \frac{\sigma_{N1}}{\sigma_{NN}}\frac{h_N^6}{\sigma_{NN}^6}
  =\sigma_{N1}h_N^6 \frac{1}{\sigma_{NN}^7}
  \leq  \sigma_{N1}h_N^6 \sum_{n=1}^N\frac{1}{\sigma_{Nn}^7}.
\end{align*}
Due to \eqref{align:hN-upper-lower-bound}, we may replace $h_N$ by $\bar h_N$, and suffer only a constant depending on $\alpha$ and $C_\pi$.  From the definition of $\bar h_N$, we may replace $\bar h_N$ with $\left( \sum_{n=1}^N \sigma_{Nn}^{-4} \right)^{-1/4}$ and suffer only a constant depending on $\alpha$.  We therefore have (with $C'$ depending only on $\alpha$ and $C_\pi$)
\begin{align*}
  \sigma_{N1}h_N^6 \sum_{n=1}^N\frac{1}{\sigma_{Nn}^7}
  &\leq  C' \sigma_{N1}\left(\sum_{n=1}^N\frac{1}{\sigma_{Nn}^7}\right)
  \left( \sum_{n=1}^N\frac{1}{\sigma_{Nn}^4} \right)^{-3/2},
\end{align*}
which tends to zero due to our assumption \eqref{align:spectral-decay-assumption}, and so too then does $h_N/\sigma_{Nn}$.

\subsection{The normal limit}
\label{section:proofs:normal-limit}
\begin{proof}[Proof of theorem \ref{theorem:integration-error}]
  Since, in equilibrium, the performance of HMC with momentum term $\|\xi\|^2/2$ is invariant under isometries (see section 4 of \citep{Neal2012-jd}), we may assume without loss of generality that our distribution is a centered diagonal Gaussian with covariance $\mbox{Diag}(\sigma_{N1}^2,\ldots,\sigma_{NN}^2)$.  Leapfrog integration will act independently on each component.  Moreover, in equilibrium, position and momentum samples from each component are independent.

  We first consider the case of one component with variance $\sigma^2$, and hide the dependence on $N$.
  Each leapfrog step is an iteration of the matrix
  \begin{align*}
    U_h :&= \left( 
    \begin{matrix}
      1 - \frac{h^2}{2\sigma^2} & h \\
      -\left( \frac{h}{\sigma^2} - \frac{h^3}{4\sigma^4} \right) & 1 - \frac{h^2}{2\sigma^2}
    \end{matrix}
    \right),
  \end{align*}
  which has eigenvalues and eigenvectors
  \begin{align*}
    \lambda_{\pm} :&= 1 - \frac{h^2}{2\sigma^2} \pm i\frac{h}{\sigma}\sqrt{1 - \frac{h^2}{4\sigma^2}},\qquad
    v_{\pm} := \left( 1, \pm \frac{i}{\sigma}\sqrt{1 - \frac{h^2}{4}} \right).
  \end{align*}
  If $h / (2\sigma) < 1$, the eigenvalues have modulus 1 and the iteration is stable.  Then, by diagonalizing, one can show
  \begin{align*}
    U_h^\ell &= \left( 
    \begin{matrix}
      \cos(\ell\theta) & \gamma^{-1}\sin(\ell\theta) \\
      -\gamma\sin(\ell\theta) & \cos(\ell\theta)
    \end{matrix}
    \right),
  \end{align*}
  where
  \begin{align*}
    \gamma :&= \sqrt{\frac{1}{\sigma^2} - \frac{h^2}{4\sigma^2}},\qquad \theta := \cos^{-1}\left( 1 - \frac{h^2}{2\sigma^2} \right).
  \end{align*}

  To compute the Hamiltonian after $\ell$ steps, we apply $U_h^\ell$ to the starting point $(x_0,\xi_0)$, then plug into $H(x,\xi) = x^2/(2\sigma^2) + \xi^2/2$ to get
  \begin{align}
    \label{align:hamiltonian-after-l-steps}
    \begin{split}
    H(U_h^\ell (x_0, \xi_0)) &=
    \cos^2(\ell\theta) \left( \frac{x_0^2}{2\sigma^2} + \frac{\xi_0^2}{2} \right)
    + \sin^2(\ell\theta) \left( \gamma^2\sigma^2 \frac{x_0^2}{2\sigma^2} + \frac{1}{\gamma^2\sigma^2}\frac{\xi_0^2}{2} \right) \\
    &+ \cos(\ell\theta)\sin(\ell\theta)\left( \frac{1}{\gamma\sigma} - \gamma\sigma \right)\frac{x_0\xi_0}{\sigma}.
    \end{split}
  \end{align}

  Define
  \begin{align*}
    \chi :&= \left( \frac{h}{2\sigma} \right)^4\cdot \frac{1}{1 - \left( \frac{h}{2\sigma} \right)^2},
  \end{align*}
  then using the relations
  \begin{align*}
    \gamma^2\sigma^2 + \frac{1}{\gamma^2\sigma^2} &= 2 + \chi,\qquad \gamma\sigma - \frac{1}{\gamma\sigma} = \sqrt{\chi},
  \end{align*}
  and the fact that the initial Hamiltonian is $x_0^2/(2\sigma^2) + \xi_0^2/2$, we find
  \begin{align}
    \label{align:delta-l}
    \begin{split}
    \delta^\ell :&=
    \frac{\sin^2(\ell\theta)}{2}\left( \frac{h}{2\sigma} \right)^2\left( \xi_0^2 - \frac{x_0^2}{\sigma^2} \right)
    + \sin^2(\ell\theta)\chi\frac{\xi_0^2}{2}
    + \cos(\ell\theta)\sin(\ell\theta)\sqrt{\chi}\frac{x_0\xi_0}{\sigma}.
    \end{split}
  \end{align}
  This has the bound
  \begin{align}
    \label{align:delta-l-bound}
    |\delta^\ell|
    &\leq \frac{h^2}{8\sigma^2}\left( \xi_0^2 - \frac{x_0^2}{\sigma^2} \right) + \chi\frac{\xi_0^2}{2} + \sqrt{\chi}\frac{|x_0\xi_0|}{\sigma}.
  \end{align}
  To compute moments, use the fact that, in equilibrium, $x_0\sim\calN(0, \sigma^2)$, and $\xi_0\sim\calN(0, 1)$ are independent.
  \begin{align*}
    \Exp{\delta^\ell} &= \frac{\sin^2(\ell\theta)}{2} \left( \frac{h}{2\sigma} \right)^4 + R_1(\delta^\ell) \left( \frac{h}{2\sigma} \right)^6,\\
    \Var{\delta^\ell} &= 2\Exp{\delta^\ell} + R_2(\delta^\ell) \left( \frac{h}{2\sigma} \right)^6,
  \end{align*}
  where there exists a constant $C<\infty$, uniform in $(\sigma,\ell,\theta)$ (so long as $h < 2\sigma$), such that $|R_j|\leq C$.
  
  Re-introducing dependence on $N$, $n$, and setting $\ell = T/h_N$ for random integration time $T\sim\pi_N$ (we assume $\ell$ is an integer, if not minor adjustments are needed), we have
  \begin{align}
    \label{align:expectation-mid-proof}
    \begin{split}
      \Exp{\Delta_N}
      &= \frac{1}{2}\sum_{n=1}^N \left( \frac{h_N}{2\sigma_{Nn}}\right)^4 \int \sin^2\left( \frac{t}{h_N}\cos^{-1}\left( 1 - \frac{h_N^2}{2\sigma_{N,n}^2} \right) \right)\pi_N(t)\dt \\
      &\quad + \frac{1}{2}\sum_{n=1}^N \left( \frac{h_N}{2\sigma_{Nn}}\right)^6 \int R_1(\delta^\ell_{N,n}) \pi_N(t)\dt.
    \end{split}
  \end{align}
  The second term is bounded in absolute value by a constant times
  \begin{align*}
    \left( \frac{h_N}{\sigma_{NN}} \right)^2 h_N^4 \sum_{n=1}^N \frac{1}{\sigma_{Nn}^4},
  \end{align*}
  which tends to zero due to \eqref{align:step-size-sigma-ratio-uniform-bound} and \eqref{align:hN-upper-lower-bound}.
  As for the first term, upon solving $1 - \eps = \cos(y)$ we have the relation $\cos^{-1}(1 - \eps) = \sqrt{2\eps} + O(\eps^{3/2})$. This means
  \begin{align*}
    \sin^2\left( \frac{t}{h_N} \cos^{-1}\left( 1 - \frac{h_N^2}{2\sigma_{N,n}^2} \right) \right)
    &= \sin^2\left( \frac{t}{h_N} \left[ \frac{h_N}{\sigma_{Nn}} + O\left( \left( \frac{h_N}{\sigma_{Nn}} \right)^3 \right) \right] \right) \\
      &= \sin^2\left( \frac{t}{\sigma_{Nn}} \right) + t\cdot O\left( \frac{h_N^2}{\sigma_{Nn}^3} \right),
  \end{align*}
  where $O\left( h_N^2/\sigma_{Nn}^3 \right)$ is a term bounded (uniformly in $N$, and $n$) by a constant times $h_N^2/\sigma_{Nn}^3$.
  Since $\int t\cdot\pi_N(t)\dt = \sigma_{N1}\Exp{T}$, we have
  \begin{align}
    \label{align:Delta-N-mean}
    \begin{split}
      \lim_{N\to\infty}\Exp{\Delta_N}
      &= \lim_{N\to\infty}
      \frac{1}{2}\sum_{n=1}^N \left( \frac{h_N}{2\sigma_{Nn}}\right)^4 \int \sin^2\left( \frac{t}{h_N}\cos^{-1}\left( 1 - \frac{h_N^2}{2\sigma_{N,n}^2} \right) \right)\pi_N(t)\dt \\
      &= \lim_{N\to\infty}
      \frac{1}{2}\sum_{n=1}^N \left( \frac{h_N}{2\sigma_{Nn}}\right)^4
      \left\{ 
      \int \sin^2\left( \frac{t}{h_N}\right)\pi_N(t)\dt
      + O\left( \frac{\sigma_{N1}h_N^2}{\sigma_{Nn}^3} \right).
      \right\} \\
      &= \frac{\alpha}{2}
      + \sum_{n=1}^N O\left( \frac{\sigma_{N1}h_N^6}{\sigma_{Nn}^7} \right).
    \end{split}
  \end{align}
  The first term is the desired limit.  The second term tends to zero upon replacement of $h_N$ by $\sum_{n=1}^N \sigma_{Nn}^{-4}$ (as in section \ref{section:proofs:preliminaries}) and the use of assumption \eqref{align:spectral-decay-assumption}.

  The steps for variance are similar, and yield
  \begin{align*}
    \lim_{N\to\infty}\Var{\Delta_N} &= \alpha.
  \end{align*}
  The normal limit follows after verifying the Lindeberg condition \citep{Durrett2010-ae}:  For all $\eps>0$,
  \begin{align}
    \label{align:lindeberg-condition}
    \lim_{N\to\infty} \sum_{n=1}^N \Var{\delta^\ell_{Nn}\st |\delta^\ell_{Nn} - \Exp{\delta^\ell_{Nn}}| > \eps} = 0.
  \end{align}
  One can check that $\delta^\ell_{Nn} - \Exp{\delta^\ell_{Nn}}$ is bounded by a term similar to \eqref{align:delta-l-bound} which tends uniformly to zero, so \eqref{align:lindeberg-condition} follows.
\end{proof}

\subsection{The simple step size}
\label{section:proofs:simple-step-size}
\begin{proof}[Proof of corollary \ref{corollary:simple-step-size}]
  The simpler step size amounts to replacing the integrals \\$\int\sin^2(t/\sigma_{Nn})\pi_N(t)\dt$ in \eqref{align:Delta-N-mean} with 1/2.
  This will be implied by sufficient decay of the remainder $R_{Nn}$ in \eqref{align:sin-integral-1}.  Indeed, $|\hat\pi(\omega)|\leq C|\omega|^{-\delta}$ implies 
  \begin{align*}
    |R_{Nn}| &\leq C \left( \frac{\sigma_{Nn}}{\sigma_{N1}} \right)^\delta,
  \end{align*}

  This means (with $\lesssim$ denoting $\leq$ up to a constant depending only on $\alpha$, $C_\pi$),
  \begin{align*}
    \left|\frac{\bar h_N^4}{h_N^4} - 1\right|
    &= \bar h_N^4 \left|\bar h_N^{-4} - h_N^{-4} \right|
    \lesssim \bar h_N^4 \sum_{n=1}^N \frac{1}{\sigma_{Nn}^4}\left( \frac{\sigma_{Nn}}{\sigma_{N1}} \right)^\delta.
  \end{align*}
  For any given $K$,
  \begin{align*}
    \bar h_N^4 \sum_{n=1}^N \frac{1}{\sigma_{Nn}^4}\left( \frac{\sigma_{Nn}}{\sigma_{N1}} \right)^\delta
    &= \bar h_N^4 \sum_{n=1}^K \frac{1}{\sigma_{Nn}^4}\left( \frac{\sigma_{Nn}}{\sigma_{N1}} \right)^\delta
    + \bar h_N^4 \sum_{n=K+1}^N \frac{1}{\sigma_{Nn}^4}\left( \frac{\sigma_{Nn}}{\sigma_{N1}} \right)^\delta \\
    &\leq \bar h_N^4\sum_{n=1}^K \frac{1}{\sigma_{Nn}^4}
    + \left( \frac{\sigma_{NK}}{\sigma_{N1}} \right)^\delta  \bar h_N^4  \sum_{n=K+1}^N \frac{1}{\sigma_{Nn}^4} \\
    &\lesssim K \frac{\bar h_N^4}{\sigma_{NK}^4} + \left( \frac{\sigma_{NK}}{\sigma_{N1}} \right)^\delta.
  \end{align*}
  As $N\to\infty$, the first term tends to zero due to \eqref{align:step-size-sigma-ratio-uniform-bound}.  The second is bounded by $r_K$ by our hypothesis.  Therefore,
  \begin{align*}
    \lim_{N\to\infty} \left|\frac{\bar h_N^4}{h_N^4} - 1\right|
    &\lesssim r_K.
  \end{align*}
  Since $K$ was arbitrary, and $r_K\to0$, we have shown $\bar h_N^4/h_N \to0$.
\end{proof}

\section*{Acknowledgements}
The authors would like to acknowledge the value of many conversations with members of the TensorFlow probability team.  In particular (in alphabetical order!)
Josh Dillon,
Matt Hoffman,
Pavel Sountsov,
and Srinivas Vasudevan.

\medskip

\bibliographystyle{imsart-number}
\bibliography{hmc_cond_citations}

\end{document}